\documentclass[prd,superscriptaddress,footinbib,twocolumn,floatfix]{revtex4}
\usepackage[dvips]{graphicx}
\usepackage{amsmath,amssymb}
\usepackage{epic}
\usepackage{eepic}

\font\ommm=dvng10

\makeatletter
\renewcommand{\@makecaption}[2]{
  \vskip\abovecaptionskip
  \sbox\@tempboxa{\small\sf #1: #2}%
  \ifdim \wd\@tempboxa >\hsize
  \small\sf #1: #2\par
  \else
    \global \@minipagefalse
    \hb@xt@\hsize{\hfil\box\@tempboxa\hfil}%
  \fi
  \vskip\belowcaptionskip}
\makeatother

\newcommand{\tr}{\operatorname{tr}}

\providecommand{\abs}[1]{\lvert#1\rvert}

\def\IC{\mathbb{C}}
\def\IZ{\mathbb{Z}}
\def\IR{\mathbb{R}}
\def\IN{\mathcal{N}}

\begin{document}
\preprint{HUTP-01/A049}
\preprint{BUHEP-01-24}
\preprint{HEP-UK-0013}    

\title{Deconstructing $(2,0)$ and Little String Theories}

\author{Nima \surname{Arkani-Hamed}}
\email{arkani@physics.harvard.edu}
\altaffiliation[Permanent Address: ]{Department of Physics, UC
  Berkeley, Berkeley CA 94720}
\author{Andrew G. Cohen}
\email{cohen@andy.bu.edu}
\altaffiliation[Permanent Address: ]{Physics Department, Boston
  University, Boston MA 02215}
\affiliation{Jefferson Physical Laboratory,
  Harvard University,
  Cambridge MA 02138}
\author{David B. Kaplan}
\email{dbkaplan@phys.washington.edu}
\affiliation{Institute for Nuclear Theory, 1550, University of
  Washington, Seattle, WA 98195-1550}
\author{Andreas Karch}
\email{karch@fresnel.harvard.edu}
\author{Lubo\v{s} Motl}
\email{motl@feynman.harvard.edu}
\affiliation{Jefferson Physical Laboratory,
  Harvard University,
  Cambridge MA 02138}

\date{October 2001}

\begin{abstract}
  We argue that two four-dimensional strongly coupled superconformal
  field theories, on the Higgs branch in certain large $N$ limits,
  become respectively $(2,0)$ theory and $(1,1)$ little string theory
  in six dimensions.  We identify the spectrum of states responsible
  for the generation of the two extra dimensions and string
  winding modes. We establish the equivalence using orbifold
  realizations of the field theories and exploiting string dualities.
  We also speculate on deconstructions of M-theory.
\end{abstract}


\maketitle

\section{Introduction}
\label{sec:introduction}

Gauge theories in greater than four dimensions are of interest to
phenomenologists and string theorists. Recently a new way of realizing
higher dimensional theories was introduced~\cite{Arkani-Hamed:2001ca}
(for related work see~\cite{Hill:2000mu}).  These models are
four-dimensional gauge theories defined on a ``theory
space''---``moose''~\cite{Georgi:1986hf} or
``quiver''~\cite{Douglas:1996sw} diagrams constructed of sites and
links---representing the gauge theory degrees of freedom.  Although
the dynamics of these models is four-dimensional at high energies,
they appear higher-dimensional below some energy scale. This energy
scale is set by the expectation values of the link fields, which higgs
the gauge groups to the diagonal subgroup, allowing the fields to
``hop'' in the extra dimensions.  In this low energy regime the theory
space becomes a picture of the extra dimensions, and the effective
action becomes that of a higher-dimensional gauge theory with
latticized extra dimensions.  In reference~\cite{Arkani-Hamed:2001ca}
both non-supersymmetric and $\IN=1$ supersymmetric models which
develop one circular extra dimension were described.  The massive
gauge multiplets from the Higgs mechanism in the four-dimensional
theory are the massive Kaluza-Klein (KK) modes of the compactified
five-dimensional gauge theory.

The supersymmetric case is particularly
interesting~\cite{Arkani-Hamed:2001ca,Csaki:2001em}.  The theory space
for this example is a circle with $N$ $SU(k)$ gauge groups, all with
gauge coupling $G$, connected by chiral superfields transforming as
bi-fundamentals under the gauge groups at adjacent sites. The extra
dimension is formed along the Higgs branch of the theory, where all
chiral multiplets have the same expectation value $\Phi \mathbf{1}$.
The parameters of the resulting effective latticized five-dimensional
gauge theory are perturbatively related to those of the
four-dimensional theory as:
\begin{equation}
  \label{eq:1}
  g_5^2 = \frac{G}{\Phi}, \, 
  2\pi R = \frac{N}{G\Phi}, \, 
  a = \frac{1}{G \Phi} = \frac{2\pi R}{N}
\end{equation}
where $a$ is the effective lattice spacing in the 5th dimension.  At
distances large compared to $a$ the resulting five-dimensional gauge
theory is both approximately five-dimensionally Lorentz invariant and
$\IN=1$ supersymmetric, with corrections proportional to positive
powers of the lattice spacing $a$.  Both of these symmetries are {\em
  larger} than the original Lorentz invariance and supersymmetry in
the four-dimensional model; in particular the supersymmetry has
effectively doubled from 4 to 8 supercharges at low energies.

This theory provides a UV completion of a five-dimensional gauge
theory by reverting to four-dimensional behavior at distances shorter
than $a$.  It is tempting to try to take $a\to 0$, while holding
$g_5^2$ and $R$ fixed---in this limit the theory would be an
interacting five-dimensional, Lorentz invariant $\IN=1$ supersymmetric
theory at {\em all} distances, providing a purely four-dimensional
definition of a truly five-dimensional theory.  However note that from
\eqref{eq:1} this limit implies that $G$, the four-dimensional
Yang-Mills coupling, must be taken to infinity.  But once $G$ gets of
order $4\pi$ the four-dimensional gauge theories are better described
in the confining phase rather than the Higgs phase. The resulting
dynamics no longer describes a five-dimensional theory.  This limits
the effective lattice spacing $a > g_5^2$ and prevents completion to a
purely higher-dimensional theory.

This analysis suggests a higher-dimensional completion might be
possible in a four-dimensional theory which remains in the Higgs phase
even at strong coupling. If instead of the asymptotically free
four-dimensional gauge theories of the previous example we had started
with superconformal theories, we might hope to take the limit $G\to
\infty$ without encountering a phase transition. This is the starting
point for our present analysis: the identification of superconformal
theory spaces which, in the Higgs phase, deconstruct extra dimensions.

We will find that the $\IN=2$ version of the previous example is such
a superconformal theory. Although the massive gauge boson spectrum
suggest a five-dimensional gauge theory, surprisingly we will find a
non-perturbative set of monopoles and dyons which fill out the
spectrum for a {\em six} dimensional theory.  Similarly we will
consider a superconformal theory space which is the triangulation of a
torus. The massive gauge boson spectrum reproduces the KK modes of a
six dimensional theory directly; remarkably the non-perturbative
spectrum exhibits winding modes!

In the continuum limit described above these theories become
well-defined interacting six dimensional Lorentz invariant theories
with 16 supercharges. Only two such theories are thought to exist: the
$(2,0)$ superconformal theory and little string theory.  The $(2,0)_k$
superconformal field theory (SCFT), defined as a $5+1$-dimensional
conformal theory with a chiral $(2,0)$ supersymmetry in six
dimensions~\cite{Berkooz:1997cq}, can be understood as a low energy
description of $k$ coincident M5-branes in M-theory or a decoupled
description of the $A_{k-1}$ singularity in type IIB string theory.
The $(2,0)$ superconformal field theory is useful in defining the
matrix model~\cite{Banks:1997vh} for M-theory on $T^4$.  Seiberg
realized that the matrix model for M-theory on $T^5$ and $T^5/\IZ_2$
must be formulated in terms of a new six-dimensional ``non-local''
quantum theory called little string theory
(LST)~\cite{Seiberg:1997zk,Aharony:1999ks}.  LST with a $(1,1)$ SUSY
can be found as a decoupled $g_s\to 0$ description of $k$ coincident
NS5-branes of type IIB string theory with $\alpha'$ kept fixed while
the T-dual $(2,0)$ LST is found on NS5-branes in type IIA string
theory.  Matrix model descriptions of these theories have also been
constructed~\cite{Aharony:1998an,Aharony:1998th}.

While many insights have been gained about the nature of these
six-dimensional theories, they remain mysterious and exhibit
surprising properties.  For instance, the holographic AdS/CFT
dual~\cite{Maldacena:1998re} of the $(2,0)_k$ SCFT for $k$ large
describes M-theory in the $AdS_7\times S^4$ geometry while the dual
description of little string theories involves a linear dilaton
background~\cite{Aharony:1998ub}.  An interesting topological term in
the low-energy derivative expansion of the equations of motion of the
$(2,0)$ theory~\cite{Ganor:1998ve} underlies the description of the
``tensionless strings''~\cite{Ganor:1996mu} that gain a nonzero
tension on the Coulomb branch.  While the $(2,0)$ SCFT exhibits
phenomena connected with tensionless strings, the little strings have
tension equal to the tension of ``big strings'', a Hagedorn density of
states, and can be thought of as toy models for some general features
of string theory while avoiding the complications of quantum gravity.

It is natural to suspect that our theories are a four-dimensional
Lagrangian description of $(2,0)$ and little string theory. We will
argue this result by realizing our four-dimensional field theories as
the $l_s\to 0$ limits of D3-branes on orbifolds. Using conventional S-
and T-dualities we will show that our continuum limit precisely
reproduces the definitions of $(2,0)$ SCFT and $(1,1)$ little string
theory. Our theories at finite effective lattice spacing offer a
discretization of $(2,0)$ and little string theory.

We will begin by describing our two superconformal theories that
perturbatively deconstruct extra dimensions.  To motivate our string
construction we will describe the expected spectra of elementary and
solitonic excitations. These states will have a standard
interpretation in our orbifold constructions.

\newsavebox{\nmoose}
\sbox{\nmoose}{%
\begin{picture}(0,0)
  \thicklines
  \put(0,0){\arc{200}{-2.21}{-1.72}}
  \put(0,100){\circle{30}}
\end{picture}}

\begin{figure}[htb]
  \centering
  \scalebox{.4}{%
    \begin{picture}(100,230)(-50,-110)
      \thicklines
      \rotatebox{22.5}{%
        \put(0,0){\usebox\nmoose}
        \put(0,0){\rotatebox{-45}{\usebox\nmoose}}
        \put(0,0){\rotatebox{-90}{\usebox\nmoose}}
        \put(0,0){\rotatebox{-135}{\usebox\nmoose}}
        \put(0,0){\rotatebox{-180}{\usebox\nmoose}}
        \put(0,0){\rotatebox{-225}{\usebox\nmoose}}
        \put(0,0){\rotatebox{-270}{\usebox\nmoose}}
}
        \put(-93.5,39.5){\circle{30}}
        \put(0,0){\scalebox{2}{\dottedline[.]{8}(-46.5,10)(-46.5,-10)}}
    \end{picture}}
  \caption{Theory space for an $SU(k)^N$ model.}
  \label{fig-1s}
\end{figure}
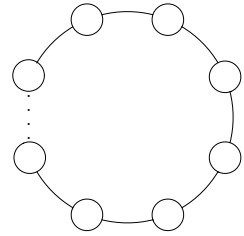

\section{Superconformal Deconstruction and Spectra}
\label{sec:superc-deconstr-spec}

Following the methods of~\cite{Arkani-Hamed:2001ca} we construct a
conformal four-dimensional $\IN=2$ gauge theory whose theory
space---that is the space of the moose/quiver diagram---is a circle
with $N$ sites, as shown in figure \ref{fig-1s}.  Each site represents
an $SU(k)$ gauge group with gauge coupling $G$, while each link
represents a hypermultiplet transforming as a bi-fundamental under the
gauge groups at the link ends.  It is a superconformal field theory,
and is expected to have an $SL(2,\IZ)$ duality~\cite{Witten:1997sc}.
This theory has a moduli space of vacua, including the direction where
each hypermultiplet has an expectation value $\Phi \mathbf{1}$.  As
in~\cite{Arkani-Hamed:2001ca} a fifth dimension is generated at low
energies, where the action takes the form of a five-dimensional
$\IN=2$ gauge theory with a latticized fifth dimension. Perturbatively
the effective lattice spacing is $a = 1/(G\Phi)$, the radius of the
new dimension is $2\pi R_5= N a=N/(G\Phi)$ and the five-dimensional
gauge coupling is $g_5^2 = G/\Phi$.  The spectrum of massive gauge
bosons is
\begin{equation}
  \label{eq:2}
  m_n^2 = 4 G^2 \Phi^2 \sin^2 \frac{\pi n}{N} = G^2\Phi^2
  \abs{e^{2\pi i n/N}-1}^2 
\end{equation}
precisely that of a latticized KK tower.
These states are not BPS. There are no BPS states
on the Higgs branch of ${\cal N}=2$ theories. So these
masses could receive quantum corrections at strong coupling. However they
are "almost BPS". That is they become BPS states in the
low energy ${\cal N}=4$ theory. Since our states have masses that put 
them in ${\cal N}=4$ up to $\frac{1}{N}$ deviations all quantum 
corrections
to the masses must go to zero as $\frac{1}{N}$. Far from the top of the
tower, $n \ll N$, the mass spectrum is nearly linear $m_n \simeq
n/R_5$, reflecting the nearly continuum physics at distances large
compared to the lattice spacing $a$. Since $\IN=2$ supersymmetry in 5
dimensions has 16 supercharges, while our four-dimensional field
theory only has 8 supercharges, we have doubled the supersymmetry at
distances large compared to the lattice spacing $a$. This is not
surprising: the effective action for a latticized fifth dimension
necessarily breaks these extra supersymmetries, but this breaking
vanishes at distances where the five-dimensional continuum description
applies.  At the very lowest energies beneath the lightest massive KK
mode, $E \ll 1/R_5$, the theory reduces to a four-dimensional $\IN=4$
$SU(k)$ gauge theory with gauge coupling $g_4^2 = G^2/N = g_5^2/(2\pi
R_5)$. As usual the compactification of a five-dimensional $\IN=2$ gauge
theory on a circle leads to a four-dimensional $\IN=4$ gauge theory
beneath the compactification scale.

In addition to the almost BPS spectrum of massive gauge bosons, this theory
has a spectrum of ``dyonic'' excitations.  We can infer the full
spectrum by exploiting the $SL(2,\IZ)$ S-duality of the $\IN=2$
conformal field theory. The strong-weak duality is generated by $G
\leftrightsquigarrow N/G$ (which interchanges $g_4^2
\leftrightsquigarrow 1/g_4^2$ in the very low energy $\IN=4$ theory).
Therefore there are magnetic states with masses
\begin{equation}
  M_n^2 = \frac{4 N^2 \Phi^2}{G^2} \sin^2 \frac{\pi n}{N} 
  \qquad n \ne  0
\end{equation}
Remarkably this is the KK spectrum for yet another dimension of radius
$2\pi R_6 =G/\Phi =g_5^2$. The S-duality can be rephrased directly as
the transformation $R_5 \leftrightsquigarrow R_6$; in this form the
existence of the 6th dimension is clear beyond the spectrum.  
The KK modes with momenta in both the fifth and sixth directions are
dyons. 

For distances large compared to $a$ our quiver theory behaves like a 6
dimensional theory with 16 supercharges. Aside from the lattice
spacing, the only breaking of conformal invariance is the
compactification.  Extrapolating these perturbative results to
$G,\Phi\to\infty$ while keeping $R_5, R_6$ fixed shrinks the lattice
spacing $a \to 0$, yielding an interacting conformal theory with the
same BPS spectrum as $(2,0)_k$ SCFT compactified on a torus. In the
next section we will show that this limit of our quiver theory {\em
  is} this $(2,0)_k$ SCFT. For finite $a \ll R_5, R_6$ our
four-dimensional quiver field theory provides a discretization of
$(2,0)$ theory. The fluctuation of each
link vev is a modulus that corresponds to a fluctuation of the
effective lattice spacing; we expect these moduli to decouple in our limit
where the lattice spacing is taken to zero \cite{Kaplan:2002wv}. (These
moduli are not apparent in the classical D-brane probe picture
presented in the next section because there, the
gauge group on the sites
is $U(k)$ rather than $SU(k)$, and the additional moduli are
lifted by the $U(1)$ D-terms). 

\begin{figure}
  \includegraphics[width=2.0in]{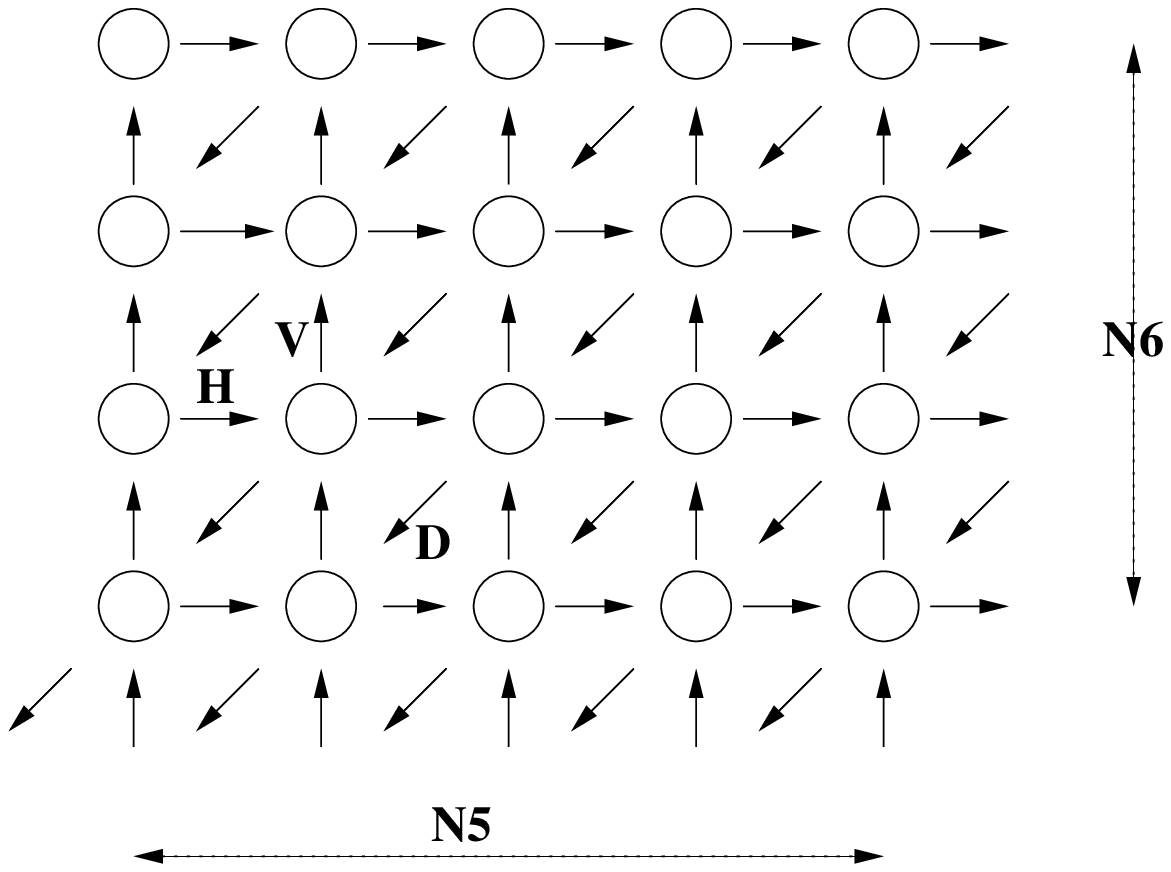}
  \caption{Theory space for the $\IZ_{N_5} \times \IZ_{N_6}$
    quiver gauge theory.}
  \label{fig:ntwoquiver}
\end{figure}

Another four-dimensional superconformal theory space is shown in
figure~\ref{fig:ntwoquiver}.  This is an $\IN=1$ gauge theory with
$N_5\times N_6$ $SU(k)$ gauge groups at the
sites~\cite{Kachru:1998ys,Lawrence:1998ja,Hanany:1998ru}.  The links
are chiral superfields transforming as bi-fundamentals under the gauge
groups at their ends. In addition we include a cubic superpotential
for each triangle: $W = \pm \sqrt{2}\lambda \tr HVD$ where the plus
sign is for clockwise oriented triangles and the minus sign for
anti-clockwise orientation. This theory has a line of superconformal
fixed points in the $(G,\lambda)$ plane~\cite{Leigh:1997ds}.
Perturbatively this begins as the straight line $\lambda = G$.  This
theory has a large moduli space of vacua, including the direction
where each $H$ and $V$ have expectation values $\Phi_5 \mathbf{1}$ and
$\Phi_6 \mathbf{1}$ respectively, while $D$ vanishes.  In this case
two extra dimensions are generated at low energies, where the action
takes the form of a six-dimensional $\IN=(1,1)$ gauge theory on a
rectangular torus with two latticized dimensions. The effective
lattice spacing is again $a_{5,6} = 1/(G\Phi_{5,6})$, the radii of the
new dimensions are $2\pi R_5 = N_5 a_5 = N_5/(G\Phi_5), 2\pi R_6 = N_6
a_6 = N_6/(G\Phi_6)$ and the six-dimensional gauge coupling is $g_6^2
= 1/(\Phi_5 \Phi_6)$.  We can make a non-rectangular torus by choosing
different directions in moduli space where all three of $H,V,D$
acquire expectation values.  The potential following from the
superpotential includes ``plaquette'' operators which produce the $\tr
F_{56}^2$ pieces of the six-dimensional gauge action.  The spectrum of
massive gauge bosons is that of a latticized toroidal KK tower with
mass scales $m_{5,6} = 1/R_{5,6}$.  Since $\IN=(1,1)$ supersymmetry in
6 dimensions has 16 supercharges, while our four-dimensional field
theory only has 4 supercharges, we have quadrupled the supersymmetry
at distances large compared to $a$.  At the very lowest energies
beneath the lightest KK mode, the theory reduces to a four-dimensional
$\IN=4$ $SU(k)$ gauge theory with gauge coupling $g_4^2 = G^2/(N_5
N_6) = g_6^2/(4\pi^2 R_5 R_6)$.

In addition to the spectrum of massive gauge bosons, this theory has a
spectrum of dyonic excitations. We can again infer the spectrum by
exploiting S-duality. Making the transformation $G
\leftrightsquigarrow N_5 N_6/G$, along with the geometric symmetry
$R_5\leftrightsquigarrow R_6$, this takes $m_{5,6} = 1/R_{5,6} \to
M_{5,6} = R_{5,6}/g_6^2$. This is a T-duality, and the dyonic spectrum
includes ``winding'' modes!

For distances large compared to $a_{5,6}$ our quiver theory behaves
like a 6 dimensional theory with 16 supercharges. In addition to the
lattice spacing, and the compactification, this theory breaks
conformal invariance through $g_6^2$.  Taking $G \to\infty$ while
keeping $R_5, R_6$ and $g_6^2$ fixed shrinks the lattice spacings
$a_{5,6} \to 0$, yielding an interacting theory with winding modes
compactified on a torus, just like the $(1,1)$ little string theory.
In the next section we will show that this limit of our quiver theory
{\em is} $(1,1)$ little string theory, with little string tension
$1/g_6^2$. Again for finite $a_{5,6} \ll R_5, R_6, g_6$ our
four-dimensional quiver field theory provides a discretization of
$(1,1)$ little string theory.

\begin{figure}
  \includegraphics[width=1.0in]{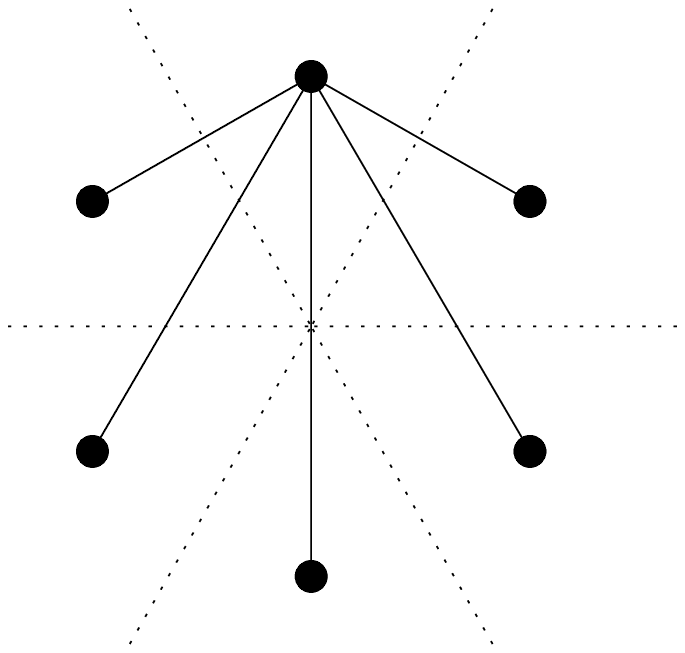}
  \caption{The W-bosons of the quiver gauge theory from
    strings stretched between D-branes and their mirror images
    under the orbifold action.}
  \label{fig:kk}
\end{figure}

\section{String Theory Analysis}
\label{sec:string-theory-analys}

\subsection{Orbifolds for Quivers}
\label{sec:orbifolds-quivers}

Because the ``latticized'' gauge theories in $d>4$ will be related to
orbifolds in string theory, it is useful to review how the quiver
gauge theories appear. Our presentation will be compressed: a complete
account can be found in the original paper by Douglas and
Moore~\cite{Douglas:1996sw}. The relevant orbifolds are $\IC^3 /
\IZ_{N}$ for the case of the $(2,0)$ superconformal field theory and
$\IC^3 / \IZ_{N_5} \times \IZ_{N_6}$ for the case of the $(1,1)$
little string theory. The former orbifold can be obtained from the
latter by setting $N_5=N$ and $N_6=1$.  The generators of $\IZ_{N_5}
\times \IZ_{N_6}$ act on three complex coordinates of $\IC^3$ as
follows:
\begin{equation}
  \label{table}
  \begin{array}{|l|l|}
    \hline \IZ_{N_5}& \IZ_{N_6}\\
    \hline z_1\to z_1 e^{+2\pi i/N_5}&  z_1\to z_1\\
    z_2\to z_2 e^{-2\pi i/N_5}&z_2\to z_2 e^{+2\pi i / N_6}\\
    z_3\to z_3& z_3\to z_3 e^{-2\pi i / N_6}\\
    \hline
  \end{array}
\end{equation}

The orbifold transformation as an action on complex coordinates is
embedded in a $U(3)$ subgroup of the possible $SO(6)$ holonomy. Since
the phases add to zero, it is actually $SU(3)$.  The orbifold is a
singular non-compact Calabi-Yau three-fold, preserving one quarter of
the original supersymmetry. For $(N_5,N_6)=(N,1)$, the orbifold action
is in $SU(2)$ and the resulting $A_{N-1}$ singularity is a non-compact
K3 manifold preserving one half of the original supersymmetry.

What is the world-volume theory of D-branes probing such an orbifold?
The gauge group and matter representations describing $k$ D-branes
will be depicted in the quiver diagram: every node of the quiver
diagram denotes a gauge group and the arrows connecting the nodes
denote the $({\bf K}, {\bf \bar K'})$ bifundamental representation of
those groups $U(K)$ and $U(K')$. In our case all $K=K'=k$ and, because
the overall $U(1)$ decouples, we will discuss $SU(k)$ groups only.

D-branes that can escape from the singularity at the origin must be
accompanied by their $\IZ_{N_5}\times \IZ_{N_6}$ mirror images on the
covering space (where we count also the original D-branes); this
number equals the order of the discrete group. If we want to describe
$k$ D-branes, it is therefore useful to start with a $U(N_5 N_6 k)$
group and project onto states invariant under the orbifold action.
The orbifold acts on the $U(N_5 N_6 k)$ Chan-Paton labels in addition
to the obvious action of (\ref{table}) on the 6 scalars
$z_i,\bar{z}_i$ and their fermionic partners.  For this action on the
group indices, we have to represent every element of $\IZ_{N_5}\times
\IZ_{N_6}$ by an element of the gauge group $U(N_5 N_6 k)$. In our
case of an abelian orbifold, the elements of the group are represented
by block diagonal matrices in the $U(k)^{N_5 N_6}$ subgroup. The
vector multiplets have to be invariant under this action and therefore
precisely the ones belonging to this $U(k)^{N_5 N_6}$ group survive.
Labeling those $N_5 N_6$ $U(k)$s by $U(k)_{l,m}$ $l=1 \dots N_5$ and
$m=1 \dots N_6$ the nodes form a lattice as in
figure~\ref{fig:ntwoquiver}.

For the scalars the geometric action has to be compensated by a
non-trivial action on the Chan-Paton factors. The scalars therefore
live in bifundamental representations.  Consider {\it e.g.} the matrix
$Z_1$ that carries the information about the positions of the D-branes
in $z_1$. The physical action of the generator of $\IZ_{N_5}$ adds a
phase $\exp(+2\pi i/N_5)$. It must be cancelled by an action on the
quiver nodes. The correct action is the shift of the nodes by one in
the horizontal direction. Therefore the field $Z_1$ transforms in the
direct sum of a fundamental representation and the anti-fundamental
representation of the group which is to the right of the first one,
one of the $H$ fields in figure~\ref{fig:ntwoquiver}. Similarly $Z_3$
carries the links $V$ in the vertical direction and $Z_2$ the diagonal
fields $D$. The fermions follow by supersymmetry and complete $H$, $V$
and $D$ into chiral multiplets.

Our superconformal field theories are obtained as the $l_s\to 0$ limit
of these orbifolds. We can match the parameters of the orbifold to the
field theory by comparing physical states. The KK modes of the field
theory correspond to fundamental strings stretching between the
D-branes and their mirror images as in figure~\ref{fig:kk}.  The
masses of these modes are the string tension times the length of the
string, reproducing the field theory KK spectrum.  The lightest KK
mode, with mass $1/R$, corresponds to the fundamental string
stretching between nearest neighbor D3-brane mirrors. The top of the
KK tower, with mass $1/a$, corresponds to the string which stretches
close to the fixed point. Similarly the magnetic and dyonic field
theory modes correspond to the D- (more generally $(p,q)$) strings
stretching between the D3-brane mirrors.

\subsection{The $(2,0)$ theory}
\label{sec:2-0-theory}

\begin{figure}
  \includegraphics[width=2.5in]{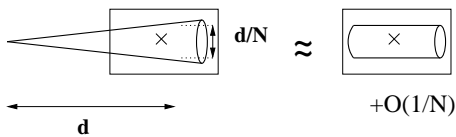}
  \caption{In the large $N$ limit the orbifold geometry locally
    degenerates into a cylinder.}
  \label{fig:cone}
\end{figure}

We wish to show that the four-dimensional $\IN=2$ circular theory of
the previous section on the Higgs branch, in the limit $N\to \infty$
holding the mass of the lightest KK mode $m$ and its dual $M$ fixed is
$(2,0)_k$ theory compactified on a torus with radii $R_5 = 1/m$ and
$R_6 = 1/M$.  We have realized this four-dimensional field theory as
the $l_s \to 0$ limit of $k$ D3-branes on a $\IC^3/\IZ_N$ orbifold.
This string theory has four parameters: $l_s, g_s, N$ and the distance
$d$ of the D3-branes from the orbifold fixed point.  The field theory
parameters are related to these by
\begin{align}
  \label{eq:3}
  \begin{split}
    m &= \frac{d}{N l_s^2}\equiv 1/R_5 \\
    M &= \frac{d}{N l_s^2 g_s} \equiv 1/R_6
  \end{split}
\end{align}
Therefore we consider this orbifold theory in the limit $l_s \to 0,
N\to \infty$ holding $g_s$ and $d/(N l_s^2)=1/R_5$ fixed. The length
of the stretched strings which correspond to the KK modes in the field
theory must be ``sub-stringy'': $d \ll l_s$.  The orbifold then
reduces to the field theory at energies $E \ll 1/l_s$.  Probing the
physics at energy $E$ involves fluctuations on the orbifold over
distances of size $\Delta x \sim E \, l_s^2$. This implies $\Delta x
/d \sim E R_5/N$.  Thus in the large $N$ limit the D3-branes only
fluctuate over distances small compared to $d$.  Consequently we may
approximate the $\IC^3/\IZ_N$ orbifold as $\IR^5 \times S^1$ where the
radius of the circle is $r = d/N \sim l_s^2/R_5$ plus corrections
which vanish as $N\to \infty$.  Figure~\ref{fig:cone} represents the
``straightening'' of the space near the D3-branes. More formally,
choosing coordinates on $\IC^3$ as $z_1= r_1 \exp(i \theta + i
\phi/N), z_2 = r_2 \exp(i\theta -i\phi/N) \text{ and } z_3$, the
orbifold metric is flat with the periodic identifications $\theta \to
\theta +2\pi, \phi\to \phi + 2\pi$. Expanding $r_1 = d/\sqrt{2} + x_1,
r_2 = d/\sqrt{2} + x_2, \theta = x_3/d$ the metric becomes
\begin{equation}
  \label{eq:4}
  ds^2 = dz_3 d\bar{z}_3 + dx_1^2 +dx_2^2 + dx_3^2 +
  \left(\frac{d}{N}\right)^2 d\phi^2 
\end{equation}
up to corrections of order $\Delta x/d \sim ER_5/N$ which vanish in
the large $N$ limit.  
In this picture the enhancement to 16
supercharges is manifest---supersymmetry breaking is associated with
the orbifold fixed point which can not probed in our limit.  

In the small $l_s$ limit the radius $r$ is sub-stringy---a better
description is obtained by taking the T-dual to turn this into a large
circle with radius $R_5$\footnote{At finite $N$, where the orbifold
  fixed point can still be probed, this T-duality involves NS5-branes
  corresponding to the orbifold fixed point~\protect\cite{Ooguri:1996wj}}.
The D3-branes become D4-branes wrapped around this large circle, while
the string coupling becomes $g_s^\prime = g_s R_5/l_s$. As $l_s$
shrinks to zero, the string coupling becomes large---a better
description is obtained by using M-theory. The wrapped D4-branes
become M5-branes wrapped on a torus with radii $R_5$ and $R_6 =
g_s^\prime l_s = g_s R_5$. The 11-dimensional Planck length is $l_p^3
= l_s^3 g_s^\prime = l_s^2 R_6$. As we take the limit $l_s \to 0$ the
11-dimensional Planck length goes to zero. This is the definition of
the $(2,0)_k$ SCFT in 6 dimensions compactified on a torus.

\subsection{Little String Theory}
\label{sec:little-string-theory}

The $\IN=1$ toroidal quiver is similar to the $\IN=2$ circular case.
We have realized this four-dimensional field theory as the $l_s \to 0$
limit of $k$ D3-branes on a $\IC^3/\IZ_{N_5}\times\IZ_{N_6}$ orbifold.
The field theory obtained from the orbifold at finite $l_s$ is not
precisely the superconformal theory described previously. The orbifold
matches to our toroidal quiver field theory with the boundary
condition $\lambda = G$ at the scale $1/l_s$. As we take the limit
$l_s\to 0$ the field theory flows to the superconformal fixed line.
The continuum limit in this case is $N_{5,6}\to \infty$ holding the
mass scales of the lightest KK modes $m_{5,6}$ and the dual winding
modes $M_{5,6}$ fixed.  This string theory has six parameters: $l_s,
g_s, N_{5,6}$ and the distances $d_{5,6} = \abs{z_{1,3}}$ of the
D3-branes from the orbifold fixed point.  The field theory parameters
are related to these by
\begin{align}
  \label{eq:5}
  \begin{split}
    m_{5,6} &= \frac{d_{5,6}}{N_{5,6} l_s^2} \equiv 1/R_{5,6} \\
    M_{5,6} &= \frac{d_{5,6}}{N_{5,6} l_s^2 g_s}
  \end{split}
\end{align}
Thus we consider the orbifold theory in the limit $l_s \to 0,
N_{5,6}\to \infty$ (with fixed $N_5/N_6$) holding $g_s$ and
$d_{5,6}/N_{5,6}$ fixed. The length of the stretched strings which
correspond to the KK modes in the field theory must be
``sub-stringy'': $d_{5,6} \ll l_s$.  The orbifold then reduces to the
field theory at energies $E \ll 1/l_s$.  Again, probing the physics at
energy $E$ involves fluctuations on the orbifold over distances of
size $\Delta x \sim E \, l_s^2$, implying $\Delta x /d_{5,6} \sim E
R_{5,6}/N_{5,6}$.  Thus in the large $N$ limit the D3-branes only
fluctuate over distances small compared to $d_{5,6}$.  Consequently we
may approximate the $\IC^3/\IZ_N$ orbifold as $\IR^4 \times T^2$ where
the radii of the $T^2$ are $r_{5,6} = d_{5,6}/N_{5,6} \sim
l_s^2/R_{5,6}$ plus corrections which vanish as $N_{5,6}\to \infty$.
In the small $l_s$ limit the radii $r_{5,6}$ are sub-stringy---a
better description is obtained by T-dualizing to turn these into two
large circles with radii $R_{5,6}$.  The D3-branes become D5-branes
wrapped around this large torus, while the string coupling becomes
$\tilde{g}_s = g_s R_5 R_6/l_s^2$. As $l_s$ shrinks to zero, this
string coupling $\tilde{g}_s$ becomes large---a better description is
obtained by S-dualizing, giving an NS5-brane wrapped on the same
torus, but with a new string coupling $g_s^\prime = 1/\tilde{g_s} =
l_s^2/(g_s R_5 R_6)$ and a new string length $l_s^\prime = l_s
\sqrt{\tilde{g_s}} = \sqrt{g_s R_5 R_6}$. As we take the limit $l_s
\to 0$ the new string coupling $g_s^\prime \to 0$ while the new string
length $l_s^\prime$ remains fixed. This is the definition of the
$(1,1)$ little string theory.

\section{Deconstructing M-theory?}

It is natural to attempt to extend our technique to deconstruction of
other non-gravitational theories. A more interesting challenge is the
deconstruction of M-theory itself. A partial realization of such a
deconstruction proceeds as follows. In type IIB string theory,
consider $k$ NS5-branes on a $\IC^2/\IZ_N$ orbifold. In the decoupling
limit $g_s \to 0$ with $l_s$ fixed, this is a $(1,0)$ little string
theory. At distances larger than $l_s$, this theory reduces to a
$(1,0)$ six-dimensional supersymmetric quiver gauge theory with the
same quiver diagram as figure 1, with $N$ additional tensor
multiplets. Along the Higgs branch of this theory, we can hope to
generate a {\em seventh} dimension of fixed radius $R_7$.  Since there
are no known non-gravitational UV completions of 7-dimensional
supersymmetric gauge theories, we expect this to result in a
gravitational theory.  The analysis mirrors our $(2,0)$ string theory
construction. We first S-dualize to obtain $k$ D5-branes on the same
orbifold, with a new string coupling $\tilde{g}_s = 1/g_s \to \infty$,
and a new string length $\tilde{l}_s= l_s \sqrt{g_s} \to 0$. As
before, the orbifold degenerates into a cylinder with radius $r \sim
\tilde{l}_s^2/R_7$ smaller than the string length. A better
description is obtained by T-dualizing to turn this into a large
circle with radius $R_7$. The D5-branes become D6-branes wrapped
around this large circle, while the string coupling becomes $g_s'=
\tilde{g}_s R_7/\tilde{l}_s$ and the string length is unchanged $l_s'
= \tilde{l}_s$.  Since $g_s' \to \infty$, a better description is
M-theory with radius $R_{11} = g_s' l_s' = R_7/g_s\to \infty$, and
11-dimensional Planck length $l_p^3 = g_s' (l_s')^3 = l_s^2 R_7$
staying fixed.  The $k$ wrapped D6-branes become an $A_{k-1}$
singularity.  The seven-dimensional gauge coupling of the theory
living on the $A_{k-1}$ singularity is $g_7^2 = R_7 l_s^2 = l_p^3$.
Therefore, the bulk 11-dimensional gravity does not decouple from the
seven-dimensional gauge dynamics. Finally we can decompactify the
seventh dimension $R_7\to \infty$ holding $g_7^2 = l_p^3 = R_7 l_s^2$
fixed.  Thus, we claim that the deconstruction of a seventh dimension
beginning with a $(1,0)_k$ little string theory describes $M$-theory
on an $A_{k-1}$ singularity with fixed 11-dimensional Planck length.
We have therefore deconstructed a fully gravitational theory using
non-gravitational little strings.  This is similar in spirit to the
approach of~\cite{Ganor:1998jx}.  It would be interesting to further
deconstruct the little strings to obtain a completely field-theoretic
description.

\section{Conclusions and Discussion}
\label{sec:concl-disc}

The exploration of four-dimensional field theories characterized by
theory spaces has revealed surprising new phenomena. In some cases,
non-gravitational extra dimensions arise from theory space at low
energies~\cite{Arkani-Hamed:2001ca}, but more generally there are new
features intrinsic to theory space itself.  This realization opens up
two complementary avenues of exploration.  In one direction, the rich
dynamics of theory space can provide new approaches to beyond the
standard model physics, such as stabilizing the electroweak
scale~\cite{Arkani-Hamed:2001nc}, lowering the scale of gauge coupling
unification~\cite{Arkani-Hamed:2001vr}, breaking
supersymmetry~\cite{Arkani-Hamed:2001ed}, and other
applications~\cite{Cheng:2001an,Csaki:2001qm,Sfetsos:2001qb,Chankowski:2001hz,Rothstein:2001tu}.
In another, extra-dimensional physics can be explored on a firm
footing in a broader context. In this paper, we have connected
deconstructed dimensions with string theory. We have argued that, with
sufficient supersymmetry, certain limits of local, {\em
  four}-dimensional supersymmetric field theories described by theory
spaces become $(2,0)$ SCFT and little string theory in {\em six}
dimensions.

It would be interesting to understand many aspects of our
correspondence in more detail. For instance, the little string theory
is known to have a Hagedorn
transition~\cite{Maldacena:1997cg,Aharony:1998tt}---how is this
reflected in the field theory? In our four-dimensional theory at the
origin of moduli space, the conformal invariance is enough to ensure
that the free-energy scales as $N_5 N_6 T^4$. Even out on the Higgs
branch where the extra dimensions are formed, this should be the
behavior of the free-energy at temperatures above the Higgsing scale,
that sets the little string tension.  Since the free-energy diverges
as $N_{5,6} \to \infty$, we might expect a limiting temperature set by
the string tension. However a direct verification of an exponential
growth of the number of states with mass set by the little string
scale would be far more convincing. Although the little strings
wrapped around the extra dimensions are likely four-dimensional
instantons ``lifted'' into the theory space, it would be nice to
construct the little strings explicitly as field theory solitons.
Doing so is important for elucidating issues such as the tension
between locality and T-duality~\cite{Seiberg:1997zk}. It would also be
interesting to understand the relation of our constructions with
matrix theory~\cite{Banks:1997vh}.

Our construction is based on orbifold realizations of our
four-dimensional field theories. In the large $N$ limit these orbifold
geometries are only probed in a way that is insensitive to the
orbifold fixed points, so that the relevant geometry is
well-approximated by compact circles cross flat space.  String
dualities can then be used to relate these field theories to known
sectors of string theory. There are likely deconstructions of other
non-gravitational theories in five and six dimensions, such as theories
with less supersymmetry, {\ommm:}~\cite{Gopakumar:2000ep} and
NCOS~\cite{Gopakumar:2000na,Seiberg:2000ms} theory, {\it etc}.
Especially interesting would be the complete deconstruction of a
gravitational theory. We have indicated a first step in the
deconstruction of M-theory, starting with a non-gravitational little
string theory in six dimensions.

\bigskip

\begin{acknowledgments} 
  A.G.C., D.B.K and N.A-H. wish to thank the Aspen Center for Physics
  where part of this work was done.  We thank
  Howard Georgi for collaboration on an early version of this work,
  and Cumrun Vafa for many discussions of the string
  realizations. We thank Jacques Distler for correcting us on 
  the discussion of
  BPS states for the ${\cal N}=2$ theory, and for other useful discussions,
  in particular about the decoupling of the local lattice fluctuations.
  We also thank Mina Aganagic,
  Michal Fabinger, Sergei
  Gukov, Shiraz Minwalla, Nathan Seiberg, and Andrew Strominger for
  useful discussions. This work was supported by the National Science
  Foundation under grant number NSF-PHY/98-02709.  A.G.C. is supported
  by the Department of Energy under grant number \#DE-FG02-91ER-40676.
  N.A-H.  is supported by the Department of Energy under
  contract DE-AC03-76SF00098, the National Science Foundation under
  grant PHY-95-14797, the Alfred P.  Sloan foundation, and the David
  and Lucille Packard Foundation.  D.B.K. is supported by the
  Department of Energy under grant number DE-FG03-00-ER-41132.  A.K.
  is supported by the NSF under grant PHY-9218167.  L.M. is supported
  by a Harvard Junior Fellowship.
\end{acknowledgments}


\providecommand{\href}[2]{#2}\begingroup\raggedright\endgroup

\end{document}